\documentclass[runningheads]{llncs}
\usepackage[T1]{fontenc}
\usepackage{graphicx,verbatim}
\usepackage{amsmath}
\usepackage{amssymb}
\usepackage{mathtools}
\usepackage{hyperref}
\usepackage{microtype}
\usepackage{subfigure}
\usepackage{booktabs} 
\usepackage{marvosym}
\usepackage{color}
\begin{document}
\title{ClinicalFMamba: Advancing Clinical Assessment using Mamba-based Multimodal Neuroimaging Fusion}
\titlerunning{Mamba-based Multimodal Neuroimaging Fusion}
%
\author{Meng Zhou\inst{1(\text{\Letter})}\thanks{Work done while the first author was at the University of Toronto.} \and Farzad Khalvati\inst{2,3}}

\authorrunning{Zhou and Khalvati}
%
\institute{TD Bank Group, Toronto, Canada \\
\and  Department of Computer Science, University of Toronto, Toronto, Canada \\
\and Department of Medical Imaging, University of Toronto, Toronto, Canada \\
\email{simonzhou@cs.toronto.edu} \\
}


\maketitle              
\begin{abstract}
Multimodal medical image fusion integrates complementary information from different imaging modalities to enhance diagnostic accuracy and treatment planning. While deep learning methods have advanced performance, existing approaches face critical limitations: Convolutional Neural Networks (CNNs) excel at local feature extraction but struggle to model global context effectively, while Transformers achieve superior long-range modeling at the cost of quadratic computational complexity, limiting clinical deployment. Recent State Space Models (SSMs) offer a promising alternative, enabling efficient long-range dependency modeling in linear time through selective scan mechanisms. Despite these advances, the extension to 3D volumetric data and the clinical validation of fused images remains underexplored. In this work, we propose ClinicalFMamba, a novel end-to-end CNN-Mamba hybrid architecture that synergistically combines local and global feature modeling for 2D and 3D images. We further design a tri-plane scanning strategy for effectively learning volumetric dependencies in 3D images. Comprehensive evaluations on three datasets demonstrate the superior fusion performance across multiple quantitative metrics while achieving \textbf{real-time fusion}. We further validate the clinical utility of our approach on downstream 2D/3D brain tumor classification tasks, achieving superior performance over baseline methods. Our method establishes a new paradigm for efficient multimodal medical image fusion suitable for real-time clinical deployment\footnote{Code will be available at \url{https://github.com/simonZhou86/clinicalfmamba}}.


\end{abstract}
\section{Introduction}

Multimodal Medical image fusion (MMIF) aggregates complementary information from multiple modalities (e.g., CT, MRI, SPECT) to produce higher-quality fused images that combine anatomical and functional details~\cite{10256252,zhou2024edge}. By integrating modality-specific strengths, MMIF reveals subtle anatomical structures and pathological features that may be missed when examining individual modalities. This enhanced visualization capability significantly improves clinical applicability, including tumor boundary localization~\cite{chen2024mr} and radiotherapy treatment planning~\cite{safari2023medfusiongan,10256252}. Due to hardware constraints and current physical imaging principles~\cite{10256252}, individual modalities can only capture specific aspects of tissue characteristics~\cite{safari2023medfusiongan}, leading to incomplete diagnostic information. MMIF addresses this challenge by integrating complementary information into a single, comprehensive image that preserves the most relevant features from each modality, thereby supporting more accurate and efficient diagnosis~\cite{wang2022functional}.

In recent years, deep learning models have significantly improved multimodal fusion performance through their powerful representation capabilities.~\cite{fu2021multiscale} proposed a CNN-based residual pyramid attention network for MRI-CT, MRI-PET, and MRI-SPECT fusion. Similarly,~\cite{li2022multiscale} introduced a double residual attention network to capture detailed features while avoiding gradient issues. However, CNN-based models remain limited by their inherent local receptive fields, which restrict their ability to capture long-range spatial dependencies. Transformer models~\cite{vaswani2017attention}, however, address CNNs' limitations in global feature extraction through self-attention mechanisms.~\cite{ma2022swinfusion} proposed SwinFusion, which combines CNN and Transformer models to capture local information while integrating global complementary features from both domains. Similarly,~\cite{10256252} proposed a multiscale CNN with residual Swin Transformer layers for effective feature learning from both domains. However, the quadratic computational complexity $O(N^2)$ of self-attention mechanisms creates prohibitive costs for clinical applications with large images, limiting the practical deployment of these methods. Recently, the improved selective structured state space models (Mamba)~\cite{gu2023mamba} provide a novel solution for long-term dependency modeling through selective scan with linear complexity. Several studies have already leveraged Mamba in medical vision tasks, including classification~\cite{yue2024medmamba}, segmentation~\cite{xing2024segmamba,li2025selective,ma2024u}, and multimodal fusion~\cite{li2024mambadfuse,xie2024fusionmamba}, achieving superior performance over CNN and Transformer counterparts. Despite these advancements, existing Mamba-based approaches primarily focus on global dependency modeling while neglecting the fine-grained local feature extraction capabilities that CNNs excel at. Current Mamba-based fusion approaches are predominantly designed for 2D images, creating a significant gap for clinical deployment where volumetric 3D data is preferred. Moreover, most fusion methods lack validation on clinical downstream tasks, limiting their real-world applicability. Therefore, developing a computationally efficient model that learns both local fine-grained and global features while achieving superior fusion performance and demonstrating clinical effectiveness remains crucial.

In this work, we propose a novel end-to-end CNN-Mamba framework for multimodal medical image fusion that effectively learns both local spatial features and global feature interactions. Our approach integrates Dilated Gated Convolution Blocks (DGCB) for multiscale feature extraction, a latent Mamba model to learn global feature interactions in 2D and 3D images, and cross-modal channel attention for feature decoding. We validate clinical applicability on high-grade glioma (HGG) versus low-grade glioma (LGG) brain tumor classification for \textbf{both 2D and 3D} settings, which is a critical task for precision diagnosis~\cite{zhou2024conditional,zhou2025generating,srinivasan2023grade}. \textbf{Our contributions are as follows: }(1). We introduce an end-to-end CNN-Mamba hybrid framework to effectively model local and global feature dependencies in 2D and 3D medical images. (2). We propose DGCB for multiscale feature learning, integrated with latent Mamba and cross-modal channel attention for cross-modal information fusion. (3). We are the first to extend the Mamba-based fusion method to 3D medical imaging through a novel tri-plane scanning strategy and provide the first benchmark evaluation of these methods on the clinical LGG/HGG brain tumor classification task. (4). Extensive experiments show our proposed method outperforms several baselines in both fusion metrics and downstream classification performance.

\section{Materials and Method}

\begin{figure}[ht] 
	\begin{center}
		\includegraphics[width=\textwidth]{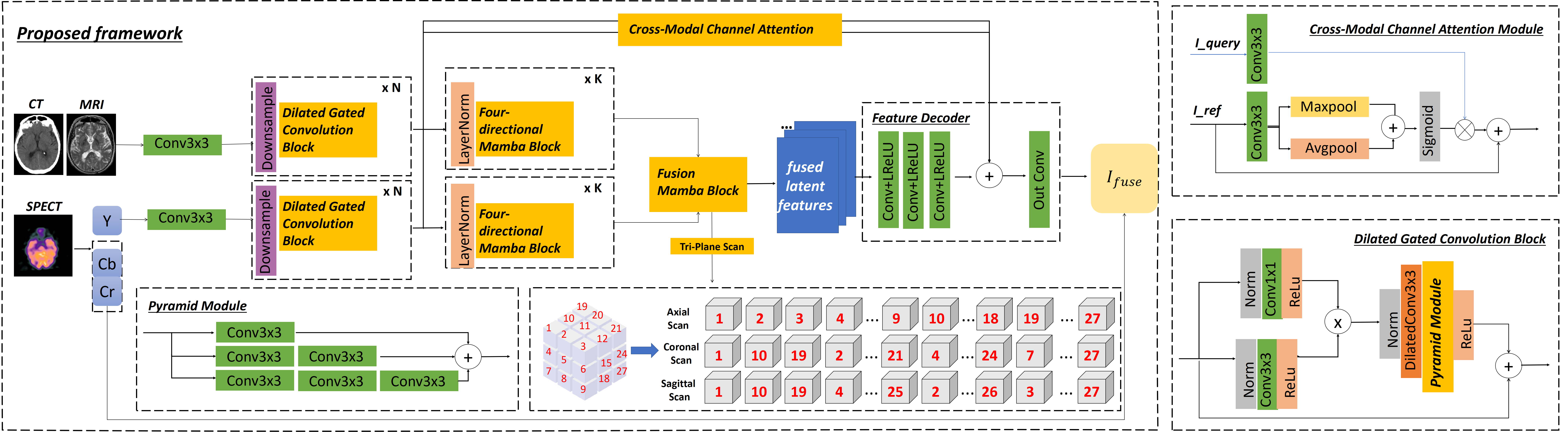}
	\end{center}
	\caption{An overview of the proposed framework. \texttt{DilatedConv3x3} represents dilated convolution with kernel size 3$\times$3. We use $N=3$ for DGCB, with dilated rate = 1,3,5, respectively, and $K=5$ for the latent Mamba model. All \texttt{Conv+LReLU} layers in the decoder have 3$\times$3 convolution followed by Leaky-ReLU. Note that the YCbCr conversion only applies to SPECT images. For 3D implementation, all operations are replaced with their corresponding 3D alternatives (i.e., Conv3$\times$3 to Conv3$\times$3$\times$3). Zoom in for a better view.}
	\label{overall}
\end{figure}

\subsection{Feature Extraction and Image Reconstruction} \label{feir}
\noindent \textbf{Hybrid Feature Encoder. }The core module in the feature encoder is the \textbf{Dilated Gated Convolution Block} (DGCB), designed to efficiently learn local spatial features. The gated mechanism enables cross-region interactions over feature maps and controls information transmission between layers, similar to~\cite{liu2021pay}. The DGCB processes input feature maps through two parallel convolution blocks with $3\times3$ and $1\times1$ kernels, respectively. We use element-wise multiplication between these features to enable cross-region interactions. Subsequently, we combine dilated convolutions~\cite{yu2015multi} and pyramid convolutions~\cite{li2018pyramid} to further capture multi-scale features and enhance discriminative feature extraction capabilities. 

\noindent \textbf{Latent Mamba for Feature Fusion. }After extracting multiscale local features, we leverage Mamba to model long-range dependencies within latent feature regions. For 2D medical images, we adopt the four-directional Mamba and Fusion Mamba blocks from~\cite{peng2024fusionmamba} to capture global contextual information across different scanning directions. However, 3D medical images possess a volumetric structure that cannot be effectively captured by conventional 2D scanning strategies. To preserve spatial coherence across all three dimensions, we propose a \textit{tri-plane} scanning strategy inspired by standard anatomical viewing planes in medical imaging. Specifically, we scan 3D feature maps along three orthogonal planes: axial, coronal, and sagittal (bottom of Figure~\ref{main_res}), enabling comprehensive structure learning from multiple views. Each plane-specific sequence is processed through Mamba layers for further processing, and the outputs are merged through Fusion Mamba blocks~\cite{peng2024fusionmamba} to form a final unified representation of the 3D volume.

\noindent \textbf{Lightweight Image Decoder. }The fused latent features are processed through convolution blocks to reconstruct the fused image. We introduce a cross-modal channel attention (CMCA) module to capture inter-channel interactions between fused and original features from both modalities. As shown in the upper right of Figure~\ref{main_res}, the cross-modal channel attention module takes two inputs $I_{ref}$ and $I_{query}$. For $I_{ref}$, we apply average pooling and max pooling to select important channel-wise representations. The combined channel map is then applied to $I_{query}$ to preserve complementary information from both modalities. There are two CMCA modules used in this work. We first use one modality as $I_{ref}$ and another modality as $I_{query}$ for the first module, then reverse the assignment for the second module. The resulting enhanced features from both operations are element-wise added to the original latent features before decoding back to image space.

\noindent \textbf{Loss Function. }Our loss function combines structural similarity (SSIM), pixel intensity, and gradient difference, following~\cite{zhou2024edge,li2024mambadfuse,xie2024fusionmamba}. Our end-to-end framework leverages a dual-target training strategy where both input modalities serve as reconstruction targets. The pixel loss is formulated as: $\mathcal{L}_{pixel} = \|\hat{x} - max(x_1, x_2)\|_{1}$; the gradient loss is: $\mathcal{L}_{grad} = \|\nabla \hat{x} - max(\nabla \hat{x_1}, \nabla \hat{x_2})\|_{2}$; and the SSIM loss is: $\mathcal{L}_{ssim} = \frac{1}{2}(1-SSIM(\hat{x}, x_1)) + \frac{1}{2}(1-SSIM(\hat{x}, x_2))$. $\hat{x}$ is the fused image, $x_1, x_2$ are the input images for two modalities, \texttt{max()} is the max operation, \texttt{SSIM()} is the structure similarity between two images. For the 3D task, we change the gradient loss to its 3D variant as introduced in~\cite{zhou2025generating} and change the \texttt{SSIM()} to \texttt{SSIM3D()}. The final loss is formulated as $\mathcal{L}(\theta) = \lambda_1 * \mathcal{L}_{pixel} + \lambda_2 * \mathcal{L}_{grad} + \lambda_3 * \mathcal{L}_{ssim}$, where we empirically set $\lambda_1 = 2, \lambda_2 = 10, \lambda_3 = 5$ for all our fusion models.

\subsection{Datasets}
In this work, we use three datasets to validate the effectiveness of our proposed approach: MRI-CT (184 pairs) and MRI-SPECT (357 pairs) multi-modality fusion data sets\footnote{\url{https://www.med.harvard.edu/aanlib/}}, and the BraTS 2019 dataset~\cite{bakas2017advancing,bakas2018identifying,menze2014multimodal} (335 patients). For MRI-SPECT fusion, we converted the SPECT images from the RGB color space to the YCbCr space following~\cite{fu2021multiscale,li2022multiscale,zhou2024edge}, using only the Y-channel images to train the model. All pairs of MRI-CT and MRI-SPECT images were coregistered and preprocessed beforehand, and we further normalized the pixel intensity to [0,1].

For the BraTS dataset, we use T2 and FLAIR sequences as done in~\cite{zhou2024edge}. We first obtained the ROIs by multiplying the images with masks, then center-cropped and reshaped the data from 240$\times$240$\times$155 to 128$\times$128$\times$128, and normalized all pixel intensities to the range of [0,1], which is used for the 3D task. For the 2D task, we further converted the data to 2D slices by slicing over the \textit{Axial plane} for each patient and only considered slices with at least 10\% non-zero pixels. 

\section{Experiments}
All programs were implemented in PyTorch. We randomly held out 30 image pairs from the MRI-CT dataset and 50 pairs from the MRI-SPECT dataset as the standalone test set. To ensure the robustness of our model, we repeated our experiments three times and ensured that we had different test sets in each run. 

To assess the usability of our fusion framework, we applied our method to a brain tumor classification task between LGG and HGG using the BraTS 2019 data. First, we randomly held out 40 \textit{patients} (20 LGG and 20 HGG patients) as a standalone test set for both 2D and 3D tasks. The remaining data are used to train both the fusion and classification models. For the classification model, we used (3D) ResNet-50 for all experiments and trained with focal loss~\cite{lin2017focal} followed by~\cite{zhou2024conditional}. We ran the classification experiment for three trials with different train-validation splits to ensure the robustness and reliability of our findings.

\noindent \textbf{Baseline Model \& Comparison.} For comparison on the 2D task, we evaluated against four state-of-the-art methods: EH-DRAN~\cite{zhou2024edge}, SwinFusion~\cite{ma2022swinfusion}, MRSCFusion~\cite{10256252} and MambaDFuse~\cite{li2024mambadfuse}. For the 3D fusion task, given the limited existing work in this domain, we implemented a 3D variant of EH-DRAN as our primary baseline comparison. For quantitative comparisons, we select five commonly used metrics in previous works~\cite{10256252,li2022multiscale,fu2021multiscale,chen2024mr,safari2023medfusiongan}: Peak signal-to-noise ratio (PSNR), Structural Similarity (SSIM)~\cite{wang2004image}, Feature Mutual Information~\cite{haghighat2011non}, Feature SSIM (FSIM)~\cite{zhang2011fsim}, and Information Entropy (EN). We assessed the performance using Area Under the Curve (AUC), F1-Score, and Accuracy for the downstream clinical validation on brain tumor classification tasks.

\section{Results and Discussions}
\subsection{Image Fusion Results}

\begin{figure}[h] 
	\begin{center}
		\includegraphics[width=\textwidth]{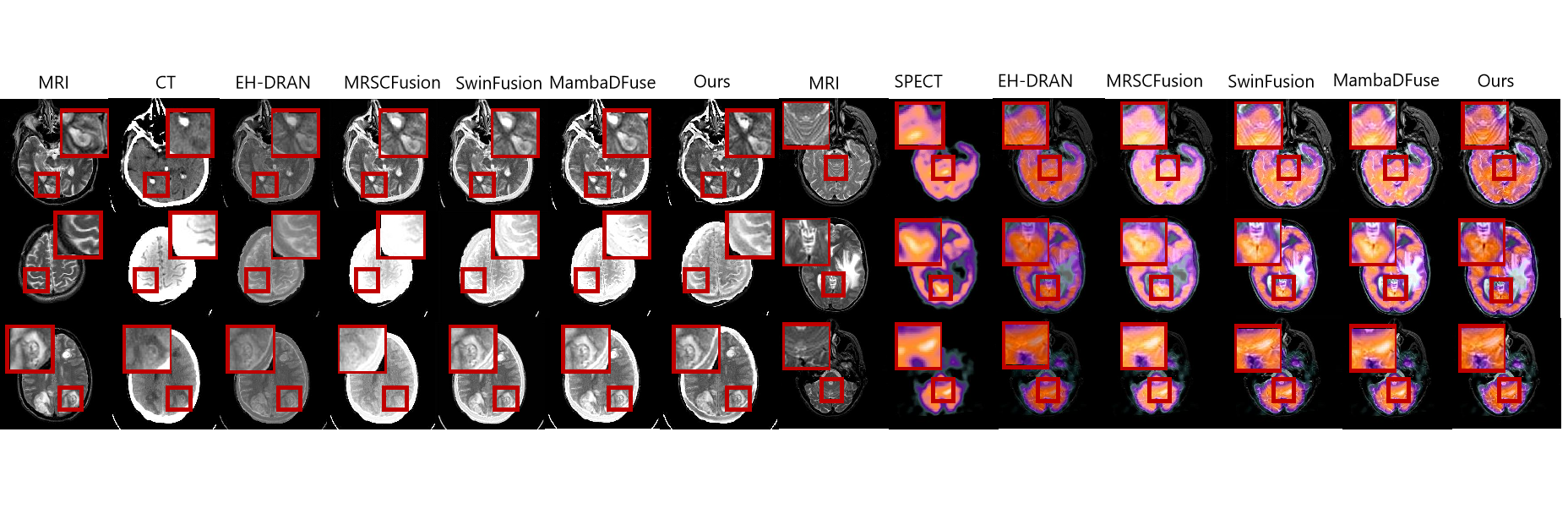}
	\end{center}
	\caption{Qualitative results for MRI-CT and MRI-SPECT fusion task. We randomly select three sample pairs from each test set and show the fusion results across different methods. Zoom in for a better view.}
	\label{qual_res}
\end{figure}

\noindent \textbf{Main Qualitative Fusion Results. }Figure~\ref{qual_res} presents qualitative comparisons of fusion results for MRI-CT and MRI-SPECT test pairs. Our qualitative analysis reveals distinct limitations across baseline methods. EH-DRAN~\cite{zhou2024edge} exhibits significant contrast degradation, producing smoothed intensity distributions that compromise both MRI details and CT structural information. MRSCFusion~\cite{10256252} generates artifacts with undesirable intensities while losing anatomical details. Although SwinFusion~\cite{ma2022swinfusion} better preserves MRI tissue characteristics, it fails to maintain adequate contrast differentiation between modalities, resulting in washed-out structural boundaries. MambaDFuse~\cite{li2024mambadfuse} suffers from substantial detail loss in MRI-derived regions and exhibits contrast distortion. Our method, however, preserves both modality-specific features and inter-modal contrast, maintaining fine-grained anatomical details from both modalities.

For MRI-SPECT fusion, similar patterns were observed: EH-DRAN~\cite{zhou2024edge} demonstrates poor contrast preservation, failing to maintain the distinctive regions present in SPECT imaging. MRSCFusion~\cite{10256252} introduces significant intensity artifacts and exhibits substantial loss of MRI texture information. While SwinFusion~\cite{ma2022swinfusion} achieves reasonable overall fusion quality, the high-intensity regions from SPECT tend to blur out fine-grained MRI details. MambaDFuse~\cite{li2024mambadfuse} shows improved functional information preservation from SPECT but continues to suffer from detail loss from MRI. Our approach achieves optimal balance, successfully preserving SPECT's functional characteristics while retaining MRI's anatomical structure and tissue contrast.

\begin{table}[h]
\caption{Comparison between different methods on two test datasets, \textbf{bold} and \underline{underline} numbers represent best and second-best values in each dataset, respectively.}
\centering
\resizebox{\columnwidth}{!}{
\begin{tabular}{ccccccc}
\hline
Dataset              & Method                                                   & PSNR$\uparrow$                      & SSIM$\uparrow$                     & FMI$\uparrow$                      & FSIM$\uparrow$                     & EN$\uparrow$                   \\ \hline
MRI-CT               & EH-DRAN            & \textbf{16.830$\pm$0.490} & 0.753$\pm$0.007    & 0.883$\pm$0.005 & \underline{0.820$\pm$0.003} & 10.727$\pm$0.531          \\
                     & SwinFusion & 14.962$\pm$0.173          & 0.768$\pm$0.007 & \underline{0.882$\pm$0.002}    & 0.810$\pm$0.001    & 8.445$\pm$0.078           \\
                     & MRSCFusion         & 14.476$\pm$0.205          & 0.713$\pm$0.012          & 0.877$\pm$0.006          & 0.791$\pm$0.010          & 7.544$\pm$0.232           \\ 
                     & MambaDFuse      &  15.873$\pm$0.289    & \underline{0.771$\pm$0.007}          & 0.882$\pm$0.005      & 0.817$\pm$0.004          & \underline{15.018$\pm$0.167}     \\
                     & ClinicalFMamba (Ours)                                            & \underline{16.519$\pm$0.352} & \textbf{0.783$\pm$0.005}    & \textbf{0.883$\pm$0.003} & \textbf{0.820$\pm$0.001} & \textbf{15.213$\pm$0.069} \\ \hline
MRI-SPECT            & EH-DRAN                                           & \underline{21.455$\pm$0.071} & 0.736$\pm$0.002 & \textbf{0.876$\pm$0.004} & \underline{0.843$\pm$0.003} & 11.970$\pm$0.538  \\
\multicolumn{1}{l}{} & SwinFusion & 17.557$\pm$0.021          & 0.728$\pm$0.004          & 0.808$\pm$0.007          & 0.819$\pm$0.011    & 13.066$\pm$0.428 \\
\multicolumn{1}{l}{} & MRSCFusion         & 18.412$\pm$0.211          & 0.734$\pm$0.012    & 0.827$\pm$0.009          & 0.814$\pm$0.006          & 9.87$\pm$0.600            \\
\multicolumn{1}{l}{} & MambaDFuse         & 21.021$\pm$0.034      & \underline{0.748$\pm$0.004} & 0.845$\pm$0.006          & 0.829$\pm$0.002          & \underline{14.126$\pm$0.439}            \\
\multicolumn{1}{l}{} & ClinicalFMamba (Ours)      & \textbf{21.561$\pm$0.067} &  \textbf{0.759$\pm$0.009} & \underline{0.856$\pm$0.003} & \textbf{0.848$\pm$0.002} & \textbf{14.871$\pm$0.334}  \\ \hline
\end{tabular}
}
\label{main_res}
\end{table}

\noindent \textbf{Quantitative Fusion Results for MRI-CT/SPECT tasks. }The quantitative metrics are shown in Table~\ref{main_res}. Our proposed method achieves the best performance in terms of SSIM, FMI, FSIM, and Information Entropy for the MRI-CT fusion task. The high FMI, FSIM, and Information Entropy scores indicate that our fused images maintain superior structural similarity and contain richer information. Our method shows a slightly lower PSNR score compared to EH-DRAN due to our method's emphasis on preserving complementary information rather than pixel-level reconstruction fidelity. For the MRI-SPECT fusion task, our method consistently surpasses baseline methods in PSNR, SSIM, FSIM, and Information Entropy. Despite a slightly lower FMI than EH-DRAN, our approach can effectively preserve more functional and morphological information from MRI and SPECT images.

\noindent \textbf{Quantitative Fusion Results on BraTS 3D images. }We evaluate our ClinicalFMamba-3D approach on the T2-FLAIR MRI fusion task to demonstrate its effectiveness on volumetric medical image fusion. Table~\ref{quan_res_3d} presents quantitative comparisons against baseline methods using PSNR, MS-SSIM~\cite{wang2003multiscale} and Entropy. Our method achieves superior performance across all evaluation criteria, which indicates better information preservation from both input modalities. These results confirm that our designed 3D Mamba effectively captures volumetric structures and enables superior fusion quality compared to its 3D-CNN counterpart.

\begin{table}[!ht]
\caption{Comparison between different methods on 3D fusion task between T2 and FLAIR images, \textbf{bold} numbers represent best performance.}
\centering
\begin{tabular}{cccc}
\hline & PSNR$\uparrow$ & MS-SSIM$\uparrow$ & EN$\uparrow$ \\
EH-DRAN-3D & 30.653$\pm$0.554 & 0.814$\pm$0.095 & 17.402$\pm$1.134 \\
ClinicalFMamba-3D (Ours) & \textbf{33.937$\pm$0.361} & \textbf{0.859$\pm$0.045} & \textbf{20.468$\pm$1.541}  \\
\hline
\end{tabular}
\label{quan_res_3d}
\end{table}

\noindent \textbf{Fusion Time Analysis. }We evaluate model efficiency by analyzing trainable parameters and fusion time on the BraTS dataset to assess clinical feasibility. Our original model contains 4.05M parameters and achieves an average fusion time of \textbf{0.1s} per image pair ($128\times128$ resolution). Our 3D variant model has 6.01M parameters with an average fusion time of \textbf{7.3s} per volume pair ($128\times128\times128$ resolution). These results demonstrate computational efficiency suitable for clinical deployment, with the 2D variant enabling real-time processing.

\noindent \textbf{Ablation Study. }We validate two key components on respective datasets. First, Cross-Modal Channel Attention (CMCA) enhances fusion by adaptively selecting informative channels across modalities. Removing CMCA module on MRI-CT data degrades performance: {\color{red}-0.552} in PSNR, {\color{red}-0.022} in SSIM, {\color{red}-0.007} in FMI, {\color{red}-0.007} in FSIM, and {\color{red}-0.592} in EN, showing the effectiveness of the proposed CMCA module. Next, we compare our proposed tri-plane scanning strategy against traditional 2D scanning (left-right, up-down only, no considerations on temporal `Z' dimension) on the BraTS dataset. Using 2D scan degrades performance: {\color{red}-7.055} in PSNR, {\color{red}-0.026} in MS-SSIM and {\color{red}-0.91} in EN, showing the superiority of our tri-plane strategy for volumetric modeling.

\subsection{Classification Results}

To validate the clinical utility of our proposed fusion framework, we conducted a downstream brain tumor classification task to distinguish between high-grade glioma (HGG) and low-grade glioma (LGG) on both 2D and 3D images. Following~\cite{zhou2024edge,xie2024mactfusion}, we utilize T2 and FLAIR sequences, as these modalities provide complementary information that can enhance classification performance when fused. We evaluated five different inputs: single-modality using T2 or FLAIR independently, dual-modality using concatenation of T2 and FLAIR; T2-FLAIR \textbf{fused image/volume} using the EH-DRAN~\cite{zhou2024edge} baseline and our ClinicalFMamba-(3D) method. The 2D classification results presented in the first block of Table~\ref{classi_res} demonstrate the superior performance of our fusion framework across all evaluation metrics. Our method achieves the highest performance with an AUC of 0.790, F1-score of 0.778, and Accuracy of 0.665, representing substantial improvements over single-modality baselines and dual-modality concatenation. Notably, our method outperforms the EH-DRAN fusion baseline by 2.1\% in AUC and 5.5\% in F1-Score, further demonstrating the effectiveness of our fusion framework. Moreover, our method maintains competitive results for the 3D classification task (second block in Table~\ref{classi_res}) with an AUC of 0.652 and an F1-score of 0.584, outperforming the EH-DRAN-3D baseline in both metrics. These results validate the capability of our fusion framework across both 2D and 3D medical imaging modalities, demonstrating strong potential for clinical usage.

\begin{table}[!ht]
\caption{Comparison of BraTS LGG/HGG 2D and 3D classification performance between different methods. Values are reported as mean$\pm$standard deviation. \textbf{bold} and \underline{underline} numbers represent best and second-best values in each dataset, respectively.}
\centering
\begin{tabular}{lcccc}
\hline
\multicolumn{1}{c}{Dataset}  & Method            & AUC$\uparrow$             & F1-Score$\uparrow$        & Accuracy$\uparrow$          \\ \hline
\multicolumn{1}{c}{BraTS-2D} & T2                & 0.722$\pm$0.021 & 0.703$\pm$0.018 & 0.604$\pm$0.037   \\
\multicolumn{1}{c}{}         & FLAIR             & 0.727$\pm$0.024 & 0.701$\pm$0.008 & 0.611$\pm$0.017   \\
\multicolumn{1}{c}{}         & T2+FLAIR          & 0.723$\pm$0.028 & 0.717$\pm$0.012 & 0.640$\pm$0.015   \\
                             & EH-DRAN           & \underline{0.769$\pm$0.003} & \underline{0.723$\pm$0.006} & \underline{0.640$\pm$0.011}   \\
                             & ClinicalFMamba (Ours)    & \textbf{0.790$\pm$0.013} & \textbf{0.778$\pm$0.023} & \textbf{0.665$\pm$0.004}   \\ \hline
\multicolumn{1}{c}{BraTS-3D} & T2-3D             & \underline{0.647$\pm$0.022} & 0.560$\pm$0.029 & 0.635$\pm$0.041   \\
                             & FLAIR-3D          & 0.641$\pm$0.110 & 0.529$\pm$0.223 & 0.566$\pm$0.010   \\
                             & T2-3D+FLAIR-3D    & 0.636$\pm$0.072 & 0.540$\pm$0.145 & 0.630$\pm$0.027   \\
                             & EH-DRAN-3D        & 0.646$\pm$0.015      & \underline{0.571$\pm$0.037}              & \textbf{0.657$\pm$0.016}               \\
                             & ClinicalFMamba-3D (Ours) & \textbf{0.652$\pm$0.038} & \textbf{0.584$\pm$0.023} & \underline{0.647$\pm$0.013}
                             \\ \hline
\end{tabular}
\label{classi_res}
\end{table}

\section{Conclusions}

\noindent We present ClinicalFMamba, a novel end-to-end CNN-Mamba hybrid architecture for multimodal 2D and 3D medical image fusion. Our framework integrates Dilated Gated Convolution Blocks for multiscale feature extraction and a latent Mamba model for cross-modal integration with long-range dependency modeling. For 3D images, we design a tri-plane scanning strategy to effectively model volumetric dependencies. Extensive evaluations demonstrate superior performance over baselines in both quantitative fusion metrics and downstream brain tumor classification, validating clinical utility with real-time processing speeds. We envision that our approach can be applied to real-world clinical settings.

    

\begin{credits}
\subsubsection{\ackname} No funding was received for conducting this study.

\subsubsection{\discintname}
The authors have no competing interests to declare that are relevant to the content of this article.
\end{credits}

%
%
%
\bibliographystyle{splncs04}
\bibliography{mybibliography}

\begin{thebibliography}{10}
\providecommand{\url}[1]{\texttt{#1}}
\providecommand{\urlprefix}{URL }
\providecommand{\doi}[1]{https://doi.org/#1}

\bibitem{bakas2017advancing}
Bakas, S., Akbari, H., Sotiras, A., Bilello, M., Rozycki, M., Kirby, J.S., Freymann, J.B., Farahani, K., Davatzikos, C.: Advancing the cancer genome atlas glioma mri collections with expert segmentation labels and radiomic features. Scientific data  \textbf{4}(1),  1--13 (2017)

\bibitem{bakas2018identifying}
Bakas, S., Reyes, M., Jakab, A., Bauer, S., Rempfler, M., Crimi, A., Shinohara, R.T., Berger, C., Ha, S.M., Rozycki, M., et~al.: Identifying the best machine learning algorithms for brain tumor segmentation, progression assessment, and overall survival prediction in the brats challenge. arXiv preprint arXiv:1811.02629  (2018)

\bibitem{chen2024mr}
Chen, W., Li, Q., Zhang, H., Sun, K., Sun, W., Jiao, Z., Ni, X.: Mr--ct image fusion method of intracranial tumors based on res2net. BMC Medical Imaging  \textbf{24}(1), ~169 (2024)

\bibitem{fu2021multiscale}
Fu, J., Li, W., Du, J., Huang, Y.: A multiscale residual pyramid attention network for medical image fusion. Biomedical Signal Processing and Control  \textbf{66},  102488 (2021)

\bibitem{gu2023mamba}
Gu, A., Dao, T.: Mamba: Linear-time sequence modeling with selective state spaces. arXiv preprint arXiv:2312.00752  (2023)

\bibitem{haghighat2011non}
Haghighat, M.B.A., Aghagolzadeh, A., Seyedarabi, H.: A non-reference image fusion metric based on mutual information of image features. Computers \& Electrical Engineering  \textbf{37}(5),  744--756 (2011)

\bibitem{li2025selective}
Li, G., Huang, Q., Wang, W., Liu, L.: Selective and multi-scale fusion mamba for medical image segmentation. Expert Systems with Applications  \textbf{261},  125518 (2025)

\bibitem{li2018pyramid}
Li, H., Xiong, P., An, J., Wang, L.: Pyramid attention network for semantic segmentation. arXiv preprint arXiv:1805.10180  (2018)

\bibitem{li2022multiscale}
Li, W., Peng, X., Fu, J., Wang, G., Huang, Y., Chao, F.: A multiscale double-branch residual attention network for anatomical--functional medical image fusion. Computers in Biology and Medicine  \textbf{141},  105005 (2022)

\bibitem{li2024mambadfuse}
Li, Z., Pan, H., Zhang, K., Wang, Y., Yu, F.: Mambadfuse: A mamba-based dual-phase model for multi-modality image fusion. arXiv preprint arXiv:2404.08406  (2024)

\bibitem{lin2017focal}
Lin, T.Y., Goyal, P., Girshick, R., He, K., Doll{\'a}r, P.: Focal loss for dense object detection. In: Proceedings of the IEEE international conference on computer vision. pp. 2980--2988 (2017)

\bibitem{liu2021pay}
Liu, H., Dai, Z., So, D., Le, Q.V.: Pay attention to mlps. Advances in neural information processing systems  \textbf{34},  9204--9215 (2021)

\bibitem{ma2022swinfusion}
Ma, J., Tang, L., Fan, F., Huang, J., Mei, X., Ma, Y.: Swinfusion: Cross-domain long-range learning for general image fusion via swin transformer. IEEE/CAA Journal of Automatica Sinica  \textbf{9}(7),  1200--1217 (2022)

\bibitem{ma2024u}
Ma, J., Li, F., Wang, B.: U-mamba: Enhancing long-range dependency for biomedical image segmentation. arXiv preprint arXiv:2401.04722  (2024)

\bibitem{menze2014multimodal}
Menze, B.H., Jakab, A., Bauer, S., Kalpathy-Cramer, J., Farahani, K., Kirby, J., Burren, Y., Porz, N., Slotboom, J., Wiest, R., et~al.: The multimodal brain tumor image segmentation benchmark (brats). IEEE transactions on medical imaging  \textbf{34}(10),  1993--2024 (2014)

\bibitem{peng2024fusionmamba}
Peng, S., Zhu, X., Deng, H., Lei, Z., Deng, L.J.: Fusionmamba: Efficient image fusion with state space model. arXiv preprint arXiv:2404.07932  (2024)

\bibitem{safari2023medfusiongan}
Safari, M., Fatemi, A., Archambault, L.: Medfusiongan: multimodal medical image fusion using an unsupervised deep generative adversarial network. BMC Medical Imaging  \textbf{23}(1), ~203 (2023)

\bibitem{srinivasan2023grade}
Srinivasan, S., Bai, P.S.M., Mathivanan, S.K., Muthukumaran, V., Babu, J.C., Vilcekova, L.: Grade classification of tumors from brain magnetic resonance images using a deep learning technique. Diagnostics  \textbf{13}(6), ~1153 (2023)

\bibitem{vaswani2017attention}
Vaswani, A., Shazeer, N., Parmar, N., Uszkoreit, J., Jones, L., Gomez, A.N., Kaiser, {\L}., Polosukhin, I.: Attention is all you need. Advances in neural information processing systems  \textbf{30} (2017)

\bibitem{wang2022functional}
Wang, G., Li, W., Gao, X., Xiao, B., Du, J.: Functional and anatomical image fusion based on gradient enhanced decomposition model. IEEE Transactions on Instrumentation and Measurement  \textbf{71},  1--14 (2022)

\bibitem{wang2004image}
Wang, Z., Bovik, A.C., Sheikh, H.R., Simoncelli, E.P.: Image quality assessment: from error visibility to structural similarity. IEEE transactions on image processing  \textbf{13}(4),  600--612 (2004)

\bibitem{wang2003multiscale}
Wang, Z., Simoncelli, E.P., Bovik, A.C.: Multiscale structural similarity for image quality assessment. In: The Thrity-Seventh Asilomar Conference on Signals, Systems \& Computers, 2003. vol.~2, pp. 1398--1402. Ieee (2003)

\bibitem{xie2024fusionmamba}
Xie, X., Cui, Y., Ieong, C.I., Tan, T., Zhang, X., Zheng, X., Yu, Z.: Fusionmamba: Dynamic feature enhancement for multimodal image fusion with mamba. arXiv preprint arXiv:2404.09498  (2024)

\bibitem{xie2024mactfusion}
Xie, X., Zhang, X., Tang, X., Zhao, J., Xiong, D., Ouyang, L., Yang, B., Zhou, H., Ling, B.W.K., Teo, K.L.: Mactfusion: Lightweight cross transformer for adaptive multimodal medical image fusion. IEEE Journal of Biomedical and Health Informatics  (2024)

\bibitem{10256252}
Xie, X., Zhang, X., Ye, S., Xiong, D., Ouyang, L., Yang, B., Zhou, H., Wan, Y.: Mrscfusion: Joint residual swin transformer and multiscale cnn for unsupervised multimodal medical image fusion. IEEE Transactions on Instrumentation and Measurement  \textbf{72},  1--17 (2023). \doi{10.1109/TIM.2023.3317470}

\bibitem{xing2024segmamba}
Xing, Z., Ye, T., Yang, Y., Liu, G., Zhu, L.: Segmamba: Long-range sequential modeling mamba for 3d medical image segmentation. In: International Conference on Medical Image Computing and Computer-Assisted Intervention. pp. 578--588. Springer (2024)

\bibitem{yu2015multi}
Yu, F., Koltun, V.: Multi-scale context aggregation by dilated convolutions. arXiv preprint arXiv:1511.07122  (2015)

\bibitem{yue2024medmamba}
Yue, Y., Li, Z.: Medmamba: Vision mamba for medical image classification. arXiv preprint arXiv:2403.03849  (2024)

\bibitem{zhang2011fsim}
Zhang, L., Zhang, L., Mou, X., Zhang, D.: Fsim: A feature similarity index for image quality assessment. IEEE transactions on Image Processing  \textbf{20}(8),  2378--2386 (2011)

\bibitem{zhou2024conditional}
Zhou, M., Khalvati, F.: Conditional generation of 3d brain tumor regions via vqgan and temporal-agnostic masked transformer. In: Medical Imaging with Deep Learning (2024)

\bibitem{zhou2025generating}
Zhou, M., Wagner, M.W., Tabori, U., Hawkins, C., Ertl-Wagner, B.B., Khalvati, F.: Generating 3d brain tumor regions in mri using vector-quantization generative adversarial networks. Computers in Biology and Medicine  \textbf{185},  109502 (2025)

\bibitem{zhou2024edge}
Zhou, M., Zhang, Y., Xu, X., Wang, J., Khalvati, F.: Edge-enhanced dilated residual attention network for multimodal medical image fusion. In: 2024 IEEE International Conference on Bioinformatics and Biomedicine (BIBM). pp. 4108--4111. IEEE (2024)

\end{thebibliography}
\end{document}